# Adoption Factors for e-Malls in the SME Sector in Saudi Arabia

**Adel A. Bahaddad**
*Faculty of Computing and IT*
*King Abdulaziz University*
*Jeddah, Kingdom of Saudi Arabia*

**Rayed AlGhamdi**
*Faculty of Computing and IT*
*King Abdulaziz University*
*Jeddah, Kingdom of Saudi Arabia*

**Salem Alkhalaf**
*College of scinece and Arts, Computer Department*
*Qassim University*
*Alrass, Kingdom of Saudi Arabia*

*Abstract*-The small and medium-sized enterprise (SME) sector represents one of the fundamental pillars in the trade field. It contributes significantly to raising the economies of countries by providing significant numbers of job opportunities, which are beneficial to directly supporting national economies. One of the most important obstacles facing this sector in the information technology era is the lack of online trading channels with consumers, which require more financial support than the their capabilities. Therefore, e-Malls might be one of the best low-cost solutions to overcome this obstacle. Also, they provide electronic platforms that include most SME requirements for sales via electronic channels as well as offer essential technical support. According to a report published in 2013 by the Saudi Arabian Monetary Agency (SAMA), the percentage of SMEs is equivalent to 90% of the total number of companies in Saudi Arabia, which is numbered at 848,500. The e-Mall is a modern idea in Saudi Arabia that requires the use of the Disunion Of Innovation (DOI) approach to diffuse e-Malls through determining companies' requirements and difficulties. Therefore, this paper focuses on the factors that help SMEs to adopt e-Malls. A quantitative questionnaire was conducted on 108 companies in Saudi Arabia to find what obstacles and requirements they face to adopt an e-Mall and focus on the factors affecting the implementation of this system, which are divided into organizational, technical and cultural factors.

## [1] INTRODUCTION

Electronic commerce (e-commerce) is one of the most critical business models on the Internet and enables people to meet many of their business needs by merely using a keyboard and mouse. Studies have recorded gradual increases in e-commerce sales in areas such as Europe, North America, and Australia ([21],[20],[26]). Therefore, many countries have improved their e-commerce frameworks and built infrastructures to face electronic commerce challenges. However, other countries, depending on their economic size and strength such as the Middle Eastern region and Saudi Arabia in particular, have not developed their electronic commerce technology as quickly as other countries.

Many e-commerce approaches concentrate on the Business to Consumer (B2C) approach such as e-retailer, service provider, e-auction, and e-Mall [29]. These approaches should be considered and their innovations diffused to other regions. In addition, the advantages, disadvantages, and requirements of these approaches should be presented to other regions to help them to understand and adopt these e-commerce technologies.

This research relates to a specific type of e-commerce – e-Mall, which is one of the types of relations between businesses and consumers. The research region is Saudi Arabia, which is one of the largest economies in Middle Eastern countries.

Considering the clear trend in global e-commerce, which reflects the economical level in developed countries, it is obvious that Saudi Arabia lags remarkably behind. In addition, the income level of SMEs, which represent most of the companies in Saudi Arabia, does not allow companies to create their own commercial sites to communicate electronically with their customers, creating difficulties in establishing e-commerce. With this situation in mind, the question was raised whether activation of the electronic market would be beneficial in increasing the global economic status of SMEs in Saudi Arabia. Is this approach suitable for adoption by Saudi Arabia? What are the difficulties and requirements that might arise? What should be done to increase both vendor confidence in the application of e-commerce?

This paper covered a literature review of material that already exists on the topic, whether these studies span regionally, Also it covered the diffusion of Innovation definition theory that is adopted in this study, and major factors that could affect e-Mall diffusion.

## [2] LITERATE REVIEW
### 2.1. Diffusion of Innovation (DOI) overview

Diffusion of innovation occurs over an undetermined period of time—from the start of a new product launch until the marketplace has been inundated—and impacts all product sales levels [40]. As a consequence, adopting e-Mall models is important because of the increase in the number of e-commerce users on the Internet [19].

In 1962, Rogers developed the concept of product life cycle to explain the different phases of how consumers accept products after they first appear on the market. They are five categories of customers. These categories change as the product moves through its life cycle as follows:

A. Innovators. These consumers represent approximately 2.5% of all customers who buy the product at the beginning of the product's life cycle.
B. Early Adopters. These consumers represent approximately 13.5% of customers and are generally accept products after the products have been "tested" by the innovators
C. Early Majority. These consumers represent approximately 34% of customers who have been prompted by the buying activities of early adopters.
D. Late Majority. These consumers represent approximately 34% of customers.
E. Laggards. These consumers represent 16% of total sales and they usually buy the product near the end of its life cycle [46].

Rogers's Theory of the Diffusion of Innovation contains four main elements: innovation, communication channels, time, and the social system.

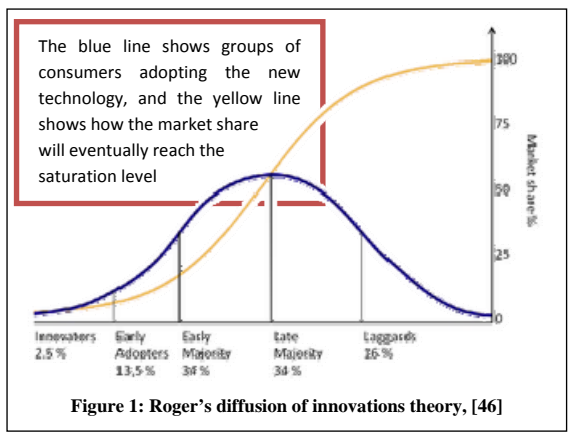

**Figure 1: Roger's diffusion of innovations theory, [46]**

A. **Innovation:** Innovation: it is an idea, practice, object, or approach that is perceived as new by a person or other unit of adoption. There are five sub-perspectives that affect the adoption of an innovation (Relative advantage – Compatibility – Complexity –Trialability –Observability).
B. Communication Channels: it can be defined as a communication relationship which can be used by contributors to generate and share information with another lead to reach a mutual understanding [16]. Another definition is the means by which messages get from one individual to another [46].
C. Time: it is the third perspective in the diffusion process. The time length for an innovation decision process for an individual begins at the first knowledge of an innovation up until that innovation is either adopted or rejected.

The innovation decision process is the process which should be executed individually or by a group to make decisions regarding any new innovation. The innovation decision process starts with the knowledge of an innovation, the crafting of an attitude regarding the innovation, making the decision to adopt or reject the innovation, and the implement and use of the new idea, and finally, confirmation of this decision [46].

D. Social System: it is defined as a culture's principles, traditional aspects, and manners that are interrelated units engaging with each other problem to solve and achieve a common goal. The participants or social units are composed of individuals, informal groups, or formal organizations [46]. The social structure influences innovation diffusion in many ways. The social system creates boundaries that effect innovation diffusion; social system can be involved to five sides. These sides are how the affects diffusion deal with system social structure, the impacts of norms on diffusions, the roles of opinion leaders and change agents, types of innovation decision, and the consequence of innovation. Each of these matters includes relationships between the social system and diffusion process that take place within it.

### 2.2. E-commerce overview

E-commerce is one of the most important online applications. The perspective of e-commerce has been more particularly addressed because of the attention it's received from both politics and the media [58]. The information explosion, facilitated by the Internet, is one of the main reasons for the spread of e-commerce. An Internet System Consortium (ISC) survey conducted in 2012 found that during the time between 2000 and June 2012, hosting names increased from 70 to more than 800 million in 245 different countries. According to Internet World State (IWS) statistics [27], global Internet users are also gradually increasing and currently represent 28% of the total population in the earth. Moreover, in the Middle East region, particularly in Saudi Arabia, the growth of Internet users has increased 4800% during the same period [27]. This clearly implies the spread of the Internet in many countries and cultures, as well as shows its impact on Internet applications, such as e-commerce [58].

At the international level, attitudes concerning e-commerce present great challenges from a policy and business perspective because e-commerce contains the possibilities to alter existing economic structures [58]. Although the first appearance of e-commerce was in the late 1990s, it was hidden within local networks and its costs were very expensive. The modern version of e-commerce, which is dependent on available frameworks, emerged and became available to all customers [29]. These frameworks, which attracted significant interest between development processes and regulatory reforms, created a technological revolution in both the economic and commercial world [58]. Also, it will be helpful to study the realignment structures of cost and depreciation, as well as application styles, channels of distribution, and after-sales service.

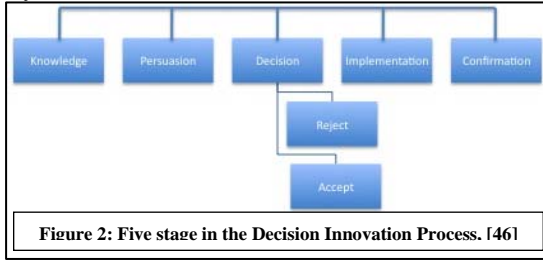

**Figure 2: Five stage in the Decision Innovation Process, [46]**

The interaction of e-commerce with other firms began in the previous decade. There are two generations of e-business applications [12]. First, the business-to-consumer (B2C) model contained simple Web pages, simple databases, and forms for purchasing and vending online products. The second generation applied e-commerce between a business and a business (B2B). This generation of e-commerce contains Web pages that are fully combined with the major legacy applications internally, along with information about business partners that is located in external systems. This provides a wide range of online services (Gonsalves, 2002).

### 2.3. Economic Growth in Saudi Arabia

In recent years, the private business sector in Saudi Arabia has grown at a rate of approximately 4.5% per annum, which some studies attribute to the emergence of many electronic systems such as e-commerce ([8],[34]). In a study which focused on the evaluation of commercial websites in Saudi Arabia [7], the most important results were as follows:

*A.* Commercial websites facilities are associated with financial firms, so 4.8 of 10 companies in Saudi Arabia have their own websites. These websites typically display information about companies, products, news and events, but about 1 in 10 sites completely support online sales.

*B.* Regarding the e-commerce difficulties in the Gulf region, the study concluded that it can be attributed to the lack of SMEs which give sufficient financial support to interactive websites that enable buying, selling and complete supervision by the company. Therefore, these companies prefer to work with third parties to compensate for the lack of expertise and qualified human resources in this sector [7]. Moreover, many businesses and consumers prefer to work with external parties for safety purposes, low costs and a broad customer base [11].

In the future, e-Malls will face three main challenges: first, how to maintain the confidence of their customers in dealing with e-Malls [3]. second, how to resolve trade disputes and deal with customer complaints [2]; and third, what support should be given to affiliate companies to increase their profits [10]. These challenges will be studied in this paper.

### 2.4. Electronic Mall (E-Mall) defined

An e-Mall can be defined as a digital environment that facilitates the interaction of consumers and retailers [57]. Zimmerman adds that e-Malls help multinational platforms communicate and conduct transactions [61]. In addition, comparisons between products' specifications and prices can be made quickly and easily ([9],[29]). Among the e-Mall frameworks, Amazon and eBay are global, Souq.com and e-Mall.com.sa are local, and E-Market and Alibaba are regional.

#### 2.4.1. Previous existing electronic Malls in Saudi Arabia

Saudi Arabia is one of broad environment in the trading exchange in the Middle East. The size of the online shopping in Saudi Arabia is estimated at around 3 SR billion ($800 million) annually, which is equivalent to about 20% of Saudi electronic trading. The electronic trading consists of bank transfers, payment methods, online shopping through electronic portals. The online trading seems is the biggest amount in the Middle East and then Egypt 2.1 billion then in Emirate 1.4 billion. Moreover, the average value of the amount paid by users for each online purchase is about 400 SR (about $107). The number of users is estimated at about 1.5 million, and the number increases 10% monthly. Additionally, about 25 million pages are visited per month [5].

SMEs are interested in entering the e-Mall field and distributing their products via the Internet for many reasons. The first is to create positive paths to grow the e-commerce sector in Saudi Arabia. Also, it is beneficial to support SMEs which do not have the capacity and efficiency of investment to build their own e-trading websites. Moreover, it spreads the Internet culture among various groups in society and increases the number of users who are interested in browsing websites as a source of interactive information and services. Furthermore, an e-Mall increases the level of technical solutions and software applications that are used to augment the effectiveness of marketing websites [51].

### 2.5. The benefits and difficulties of integrating e-Mall applications

Building strong relationships with consumers is the most important competitive advantage in the e-commerce sector. Customers can easily and quickly move quickly from one website to another. Therefore, companies should take advantage of and consider the benefits of e-Malls and any new methods to ensure customer loyalty [4]. There are many benefits and obstacles to e-Malls, as described in this paper.

#### 2.5.1 The benefit

Some studies have shown significantly higher profits from the e-Mall Model. For example, eBay has grossed nearly $3 billion in 2010. Many studies argued the e-Mall model's benefits which are focused on risk reduction currently limited to transactions via third parties, the diffusion of SMEs' products globally, improved efficiency, and reduced prices ([29],[32],[57]). So, the director of e-commerce at Microsoft agrees that the good news is sharply lower prices because of increasing competition [43]. Therefore, there are many other benefits that should be presented, which can be divided into benefits for sellers, benefits for buyers, and benefits for the purchasing process.

**2.5.1.1- E-Mall benefits**
The benefits of e-Malls to sellers are many. They can be summarized in the following points:
A. *Displaying products* through the e-Mall framework gates, which helps customers to decide whether to buy the product [12].
B. *E-Malls reduce the costs* to be paid in the traditional trade case. These costs consist of employment wages, rental shops, and advertisements. Pay for employees is a significant factor that can reduce costs in the online trading field ([12],[4],[1],[20]). Therefore, they do not need to bring large labour forces into the firm process and the practice of selling to customers in the company's branches.
C. *Online promotions* offer encouragement policies that are received by all members of the e-Mall's mailing lists. At the same time, this leads to equality when disrepute the online promotions [12].
D. *E-Malls can sell twenty-four hours a day, seven days* a week without having to pay for overtime [4].
E. *Decreasing expenditures leads to increased profit margins on goods and services* [4]. This leads to enter new segment of customers that was previously unable to purchase due to rising product costs.
F. *Wide access means* that e-Malls have the ability to deal with large segments of customers which the companies could not access them without the e-market framework ([20],[9]).
G. *E-Malls can create specific markets*, as they have the ability to create highly specific e-Malls [12].

**2.5.2 Obstacles due to regulations**
E-Malls should adopt regulations and legislation system to improve the diffusion of e-markets. Obstacles to regulations and legislation are as follows:
A. Judicial legislation refers to legislation to protect customers; it is needed for both buyers and sellers in the event of fraud or deception about a specific product or service.
B. There is a lack of specific standard tools and systems in e-Malls; these should be provided and based on the society, culture, and tradition [4].
C. The bureaucratic system is one of the most important obstacles in Saudi Arabia, and this means that approvals can take a long time to come to light [34].
D. Applicable law: Transactions are concluded electronically between the supplier and consumer, and they have a great distance between them. This may raise several legal concerns such as which laws are applicable in cases of conflict, the protection of trademarks, and the language to be adopted, especially when dealing with customers in other countries.

**2.5.3 Obstacles due to infrastructure**
E-Malls has many of the important infrastructure aspects, which are obstacles to adopting this framework in Saudi Arabia. These aspects that should be provided are as follows:
A. *Security aspect.* When buyers deal with e-Mall applications, they present credit card numbers or some personal data that could be stolen by professional website hackers ([12],[56]).
B. *Post-infrastructure in Saudi Arabia.* Unfortunately, the current service does not qualify for access to online trading, because the delivery system lacks accuracy. Moreover, the delivery system is not able to follow-up on packages, and there is no guarantee the package will arrive on time or at all [4].
C. *Delivery companies* such as DHL and FedEx. When buyers purchase goods through the Internet, delivery should be a primary focus, especially if the buyer is in a hurry. In this case the buyers have to come to their branch to receive their goods themselves which means the Delivery companies in Saudi Arabia are inefficient.
D. *Payment method* is one of the most important tools of online trading, and options are often only credit cards or a direct payment system named "SADAD." In the credit card system, there are many fees, such as for the cards themselves, service charges, interest, and so on. The large number of charges is the main reason people do not own credit cards [4]. The SADAD payment system is acceptable only in Saudi Arabia, not in other countries. In addition, it is limited by the banks and the companies that have contracts with them [47].
E. *The Internet* is the main factor that activates online trading. Although the number of Internet users is nearly 13 million [14], there are weaknesses in the infrastructure. Shortages have disrupted service for days without cause, and service is sometimes slow, especially at peak times [2]. Additionally, another issue is providing some of the suburbs and some cities with Internet services, because residents in these areas are unable to use the service and participate in e-Mall activities [1].

**2.5.4 Obstacles due to social culture and traditions**
Many of the obstacles may appear due to customs and cultural traditions in Saudi Arabia, which may affect the adoption of e-Malls there. These difficulties can be summarized as follows:
A. *Language interface:* The English language is the language of a large proportion of web sites; Major of sites are created in English language. This is a significant impediment to the citizens of many countries who are non-English speakers.
B. *Consumer behaviour:* Consumer behaviour is important because of some consumers do not trust sellers who do not see them, and they prefer to inspect the item and touch it before the completion of the procurement process.
C. *Lack of consumers' confidence:* Some researchers believe that the target audiences will not be confident enough to use online buying and selling unless these issues are solved. For example, electronic documents are not accurate and valid and cannot be accepted as official documents [4]. Moreover, consumers do not trust that their financial information transmitted via the Internet will be kept confidential [56]. All these examples represent the lack of confidence due to the newness of this technology in Saudi Arabia in particularly.
D. *There is a lack of awareness* in the persons who are involved in online trading. The lack of persons,

entities, and educational programs to help facilitate buying and selling across the Internet ([4],[53]).

The activation of those roles would be in the best economic interest of the country, companies, and individuals. Collaborative work between the services sectors would bring e-commerce to light, which would bring its advantages to the Saudi market.

### 2.6. Major factors that could affect electronic market diffusion

There are many issues associated with the adoption and application of online trading applications in organisations [33]. However, many studies and theories have been presented to study these issues and learn the appropriate frameworks to address the imbalance and shortage aspects. The factors that influence the system's performance can be devided into the technology effects related to system design, administration effects related to the structure and management of the organisation, and traditional and cultural which are related to societal requirements. In this case the factors influence can be separated into the same group which should be should be proven ([4],[2],[53]). The main purpose of this study is to determine the effect of each factor on the SMEs sector and learn about the global challenges of this study.

#### 2.6.1. People and organisational factors

These factors related to people and organisations that help to improve individual and organisational performance. These factors can be further divided into the following factors:

A. *Lack of interest and awareness:* A variety studies have been done in the online trading field ([30],[49],[53],[54]), finding that SMEs are generally unaware that their business operations can be enhanced and supported electronically by information technology perspective [12].
B. *Educational background:* Studies have shown belong ([35],[60]), which indicates that important decisions (e.g., the company's decision to adopt technology) are affected by the executive chairman's level of education.
C. *Previous IT experience:* SMEs who have previous experience using Internet technology for the same e-commerce activities are more likely to adopt a similar technique again [19],[35]).
D. *Lack of resources and IT skills:* Several studies have confirmed that a lack of resources and information technology skills is one of the key factors that hinder SMEs in the application of electronic systems and result in delayed implementation ([53],[18],[24]).
E. *Firm size:* The company's size represents one of main factors influencing the decision to adopt and use online trading in the SMEs ([23],[38]). This factor involves the volume of work, depending on the number of staff, the age of the firm, the business facilities sector such as service, manufacturing, or retail, the target population such as regional, local, national, or international, the level of technological expertise available, and the company's annual sales [12].
F. *Business sector:* There is study has shown that service-oriented companies are more likely to accept online trading companies than manufacturing firms [35]. It is thus important to consider what the most popular activities are that can be adopted via Internet and can be diffused in the Saudi Arabia's online marketing.
G. *Electronic facilitator:* It is useful to apply the e-government principle in the SMEs, even if they may have been necessary during the companies' infancy, because it allows them to replace paperwork with electronic work [36].

#### 2.6.2. Technology and environmental factors

These factors are related to the infrastructure that should be provided in e-Malls. The availability of such factors leads to an increase in the e-Mall's performance and customer confidence.

A. *Telecommunication infrastructure:* The statistics showed that more than 88% of companies in Saudi Arabia do not have the broadband service, which is available now in major cities in Saudi Arabia [14]. This issue may reduce the SMEs' performance in accessing the Internet because the other option is dial-up, which has weak Internet data exchange due to the lack of cooperation in dealing with online trading [53].
B. *Website security:* Security is a major concern for systems and organisations when trying to publish applications on the Internet. One of the key aspects of Web services management is to ensure that services can be provided and accessed according to the clear security policies and does not constitute a significant danger to the organisations during implementation [12].
C. *Delivery system:* The delivery service in Saudi Arabia is still not very effective because the government is responsible for the service, and the development remains quite slow. Currently, the postal services are not being activated to deliver mail directly to homes or offices, but there is plan to improve the infrastructure to change this situation [2].
D. *Safety payment method:* Secure payment capabilities are a key aspect of commercial transactions over the Internet. Many studies indicate that the security and speed of payment methods are important, but businesses must also be aware the payment component's infrastructure via Internet will be neither inexpensive nor simple [17]. For example of these payment methods, the credit cards such as Visa, Master Card, and American Express. However, the credit cards have not been widely used as a payment means in Saudi Arabia for three reasons. Firstly, Islam forbids people to pay an additional amount for any payable means such as credit cards. Secondly, most consumers use credit cards to withdraw cash, not to utilize the credit card's actual advantages. Thirdly, the conflicts and dispute issues about the credit card in Saudi Arabia because of the customer will usually pay the amount first, and then investigate the dispute in the hope that the issue will be solved [2]. This has led to a lack of awareness about credit card advantages, which should be understood to fulfil the main purpose of the credit card main system.

*2.6.3. Traditional and cultural factors*

The successful online trading applications in a particular region may not necessarily mean that they will be successful in another region. This is due to a number of influences, and culture and traditions represent a key factor [15]. Therefore, many of the cultural aspects are included here and cannot be overlooked.

A. *Activate inexpensive payment means cooperating:* Several studies indicate that the security and speed of payment transaction is important, however the companies should be familiar with online payment processes to be inexpensive. There are many commissions that going to be taken to accomplish the process, such as credit card fees and exchange fees from a currency to another, which affected negatively on the increasing of the cost of e-payment ([2],[11],[17]).

B. *Arabic language supported:* The Arabic language is important in order to understand many aspects that would be incomprehensible if they used a site supported by another language [3].

C. *Arabic social structure:* The e-commerce in Saudi Arabia is associated with its global counterpart, and with the country's level of economic development, technical, and social traditions. These paradoxical factors play a significant part in encouraging the promotion of e-commerce. The conservative social environment that limits women's freedom to visit Mall centres alone may positively impact the growth of e-Malls in Saudi Arabia [4].

D. *The level of development between large and small cities:* The level of development between the larger, metropolitan cities and rural areas is radically different. These significant variations could point to the positive side of e-market growth, and enhance consumers' sophistication in areas where there are fewer commercial Malls [4].

E. *Constraints that affect the application of e-Mall framework technology:* There are many difficulties should be considered when applying e-Malls in Saudi Arabia. These constraints may adversely affect the level of online trading diffusion, especially in the largest cities [4]. The as follow:
- Displaying and touching goods before the procurement process.
- Many Saudi families are interested in going Mall centres to picnic, window shop, and enhance the social side of family life.
- The customers are unable to preview and touch the products before purchasing them.
- There is a lack of counselling and advice, and lack of direct contact between suppliers and customers.
- The consumer will use invisible exchanges of money via credit cards may be a significant challenge to human nature because people tend to fear change [4].

F. *Online trading advantage:* Many online trading applications may be influenced by new facilities that are implemented with these applications. These facilities are helpful in terms of increasing the SMEs' efficiency, as well as their capacity to expand. Therefore, these features comprise a key factor for SMEs to subscribe in the e-markets, and then to activate and participate in the e-markets' frameworks ([12],[31]).

G. *The importance of e-Mall framework system effectiveness in Saudi Arabia:* A number of business organizations that use online trading systems is on the rise, but still below the expectation number (1:10) in Saudi Arabia [10]. In the future, it is likely that such systems are not only a tool to increase income, but an essential means to increase the attracting customer's number broadly [3]. Therefore, providing the personal conviction of companies' board is important step to increase profit, and Keep up with new technology to solve existing issues in business organizations

H. *Applying an innovation:* The e-Mall framework is a model for collaborative e-Mall offering a higher degree of innovation in e-commerce business models ([59],[41],[46]). As a result, it was noted that SMEs are less confident to adopt technology that has not proven successful elsewhere, especially in other countries [30].

*2.6.4. Hypothesis will assist to diffuse the e-Malls*

The fourth part consists of some of the hypotheses that depending on the consumers' conviction experienced and these hypotheses are effectiveness to diffuse the e-Mall framework in Saudi Arabia. These assumptions should be tested to gauge the importance of applying them in the e-Mall frameworks. Therefore, they are developed in this section for get statistics conclusion of the participants' opinion about these hypotheses. The hypotheses are four which are:

1. The importance to deal with e-commerce for SMEs in Saudi Arabia.
2. The importance to provide an e-Mall framework for the portal for dealers who cannot create private Internet sites.
3. The importance to Regulations and laws to protect the consumer and the seller over the Internet.
4. The e-Malls system is provide employment opportunities and reduce the spread of unemployment.

These points are intended to get target population views on these assumptions where they are important to adopt of e-Malls, but have not been tested adequately in many previous studies ([3], [4], [10]).

**2.7. Summary of Literate review**

Academic studies show that the development of e-commerce is a public demand, which these studies should be more specialised for helping to draw clear plans to enhance the economy and further define society's needs in relation to the business community. Therefore, to diffuse e-commerce in Saudi Arabia, internal aspects such as e-auctions, e-retailers and e-markets need to be generated. In addition, new sub-systems of e-commerce should be disseminated as one of the basic requirements to activate e-commerce more broadly.

Many previous studies have been published on the activation of e-commerce in Saudi Arabia, but these have indicated there has not been enough research to create e-commerce sub-systems such as e-Malls professionally. There have also been many studies on adopting global e-markets, but these have not yet been implemented to the required level in Saudi Arabia. However, it is important to note that e-commerce did not reach the Saudi community as a culture until 2001, when the Internet was established, it was available only for less than 5% of the population [14]. As end of 2013, use of the Internet is widespread, and more than 47.6% of the population uses the Internet [14]. This has assisted people to become forward thinking in terms of the beneficial uses of the Internet, such as e-commerce and e-government. In addition, researchers have been thinking about the bodies that will help to activate e-commerce in Saudi Arabia. These are divided into the commercial sector, government entities and the end consumer.

Buyers and sellers represent the key partners in the business process in B2C [29]. Therefore, the present study will focus on seller segment to assess their views when it comes to adopting e-Malls. This will illustrate whether they are convinced by this idea or whether they have reservations. Additionally, their requirements in terms of increasing their confidence in the approach will be assessed, along with difficulties that may hinder their participation and helpful methods of dealing with e-Malls in the future. Thus, the research will focus on demonstrating the difficulties and requirements related to e-Malls based on the views of the participants.

The above concerns were covered in the literature review chapter on previous research. However, there is a gap in the research: The previous literature has concentrated on activating the e-Malls, but it is also important to look at the views of the target population and what they hope to gain from this approach. These issues must be taken into account in order to accomplish the diffusion of e-Malls in Saudi Arabia to a satisfactory level.

### [3] METHODOLOGY

During collection of data relating to the adoption of e-Malls, several steps were taken to determine the appropriate way to implement e-Malls in Saudi Arabia. The research question is what factors assist traders in presenting their products via the Internet. Moreover, all e-Malls do not necessarily use the frameworks deployed in the Arabic Gulf region, which leads to different opinions about possible solutions to meet online customers' needs. Therefore, a questionnaire was selected to figure out the requirements and challenges of traders in increasing the number of online consumers. As well, this format is the best option for participants to express their views and suggestions ([4],[53]). The survey was translated into Arabic to make it easier for participants invited to fill out the survey.

| Table 1: Summary of Major factors that could affect e-Mall diffusion | | | | |
|---|---|---|---|---|
| No | Factor | | Type | Source |
| V1 | Educational background | | People & organisational factors | [35], [60] |
| V2 | Previous IT experience | | | [19], [35] |
| V3 | Firm size | | | [12], [23], [35], [38]) |
| V4 | Lack of interest and awareness | | | [30], [49], [53], [54], [12] |
| V5 | Business sector | | | [35] |
| V6 | Lack of resources and IT skills | | | [54], [18], [24] |
| V7 | Electronic facilitator | | Technology & environmental factors | [36] |
| V8 | Telecommunication infrastructure | | | [53] |
| V9 | Delivery system | | | [2] |
| V10 | Website security | | | [12] |
| V11 | Safety payment method | | | [17] |
| V12 | Activate inexpensive payment means cooperating | | Traditional & cultural factors | [2], [11], [17] |
| V13 | Arabic language supported | | | [3] |
| V14 | Families are interested with window shopping | | | [4] |
| V15 | The level of development between large and small cities | | | [3], [4], [11] |
| V16 | Arabic social structure | | | [4] |
| V17 | lack and limit direct negotiable between suppliers and customers | | | [4], [11]. |
| V18 | Online trading advantage | | | [12], [31] |
| V19 | The importance of e-Mall framework system effectiveness in Saudi Arabia | | | [3] |
| V20 | Applying an innovation | | | [41], [59], [30] |
| I1 | The importance of | Dealing with e-commerce for SMEs in Saudi Arabia | Hypotheses | [3], [4], [10] |
| I2 | | Providing an e-Mall framework for the portal for dealers who cannot own and manage their websites | | |
| I3 | | Providing Regulations and laws to protect the online customers. | | |
| I4 | | of e-Malls system is provide employment opportunities and reduce joblessness | | |

### 3.1. Research Paradigms
According to diffusion of innovation theory (DOI), important aspects should be identified and taken into account when dealing with this framework in Saudi Arabia. These aspects set essential boundaries that should be focused on and examined in a new field. These aspects were chosen when designing the questionnaire based on previous scientific studies. The aspects were identified according to five fundamental factors contained in DOI theory described earlier (See Figure 3).

### 3.2. Survey Design
The questionnaire was divided into four major parts:
A. Demographic information, covered in 11 questions
B. Factors that influence the adoption of e-Malls. These 20 factors are measured by a Likert five points scale to determine out the level of importance of each factor.
C. Open questions to learn more about factors affecting e-Malls not mentioned previously
D. Factors identified but not confirmed as influential in previous studies. These were presented as hypotheses to find out participants' opinions.

### 3.3. Choosing target participation
Some criteria were used to choose sellers as participants, as follows:
- Companies that have a CEO but have not yet adopted Internet applications
- Companies that have or plan to create a website
- Companies that plan to increase their consumers via an online trading application
- Small and medium companies with no more than 100 employees
- Companies that work in Saudi Arabia or have as their target audience Gulf populations

### 3.4. Survey distribution
The questionnaire was distributed to 140 SME companies which have branches in different cities in order to sample different opinions, cultures and traditions in Saudi Arabian society. The participants were selected through the snowballing technique. Responses were received from 85. Invitations were also sent via email to companies registered in the Jeddah Economic Gateway, a portal that has information about the activities of existing business in Jeddah, one of the largest commercial cities in Saudi Arabia [25]. More than 400 invitations were sent, but the response was very weak because most of these companies do not follow up on electronic means for communication. The participation rate was 6.25%. The distribution of the participating companies' business activities are presented in the following table.

### 3.5. Measuring Sampling size
The sample size was calculated based on the number of companies in the retail sector in Saudi Arabia: 848,500, according to the Saudi Arabian Monetary Agency, SMEs accounted for 90% [52], or an estimated 763,650 of these companies. The required sample size for this sector was 97, for a 10% margin of error and 95% confidence level. The actual number of this sample was 110.

## [4] ANALYSIS

### 4.1. The main findings for the demographic seller segment
The level of a company's performance is characterized by its ability to keep pace with global developments and build communication channels with different segments of customers, whether Locally or even internationally. These characteristics should be looked at as a way to study the company's readiness for applying new technological frameworks in new fields. Saudi Arabia is a developing country, but the successful requirements should be understood for any new framework that might be adopted. Therefore, the questionnaire aims at doing just this, and it was conducted across a wide variety of commercial activities, as found in Q1 in Table 2. This section contains questions that focus on the features of companies and their readiness to implement the technical aspects of the e-Mall system. The questions can be divided into two main parts, dealing, on one hand, with the interest of companies in applying the technical framework, and, on the other hand, with their interest in using electronic communication networks to communicate with customers via the Internet.

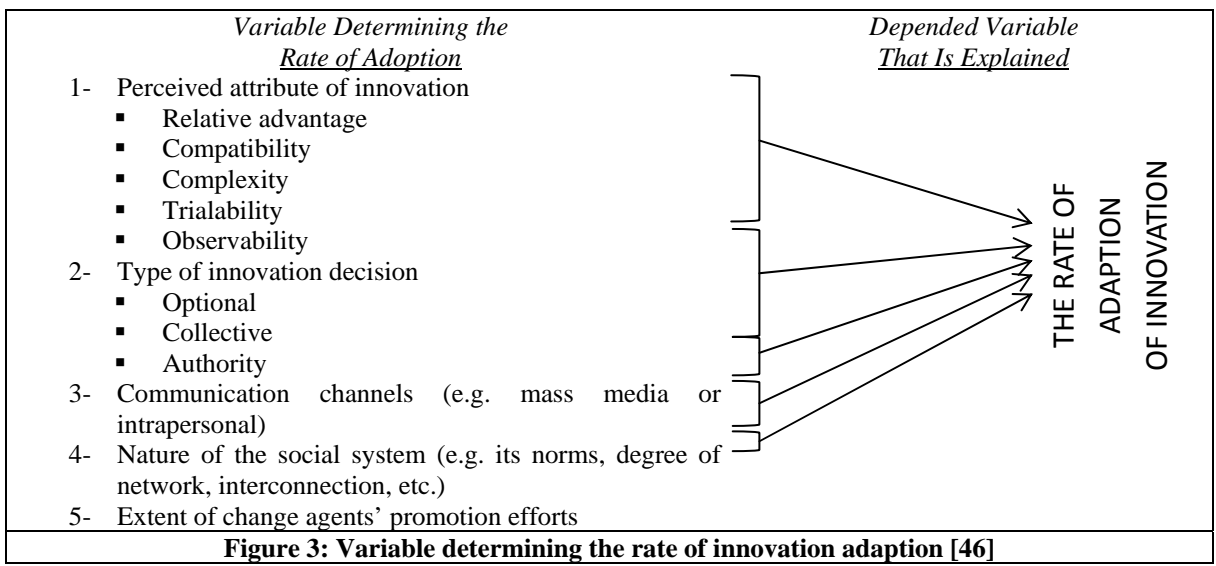

*Variable Determining the Rate of Adoption*

1- Perceived attribute of innovation
  - Relative advantage
  - Compatibility
  - Complexity
  - Trialability
  - Observability
2- Type of innovation decision
  - Optional
  - Collective
  - Authority
3- Communication channels (e.g. mass media or intrapersonal)
4- Nature of the social system (e.g. its norms, degree of network, interconnection, etc.)
5- Extent of change agents' promotion efforts

*Depended Variable That Is Explained*

THE RATE OF ADAPTION OF INNOVATION

**Figure 3: Variable determining the rate of innovation adaption [46]**

The main findings and significant points have been presented in the following sections.

**Table 2: Participating companies, broken down by type of activity**

|  | F | P |
|---|---|---|
| Real estate and public services | 6 | 5.5 |
| Books - Learning Resources | 6 | 5.5 |
| Electronics and Electrical | 18 | 16.4 |
| Car parts and accessories | 10 | 9.1 |
| Beauty and Health | 10 | 9.1 |
| Clothing | 14 | 12.7 |
| Furniture | 10 | 9.1 |
| Jewellery | 8 | 7.3 |
| Medicines and medical supplies | 2 | 1.8 |
| Household items, tools, and toys | 8 | 7.3 |
| Tickets, hotel reservations, and car rental | 6 | 5.5 |
| Programs, computer games, and Internet site | 6 | 5.5 |
| Food | 6 | 5.5 |
| Total | 110 | 100.0 |

The inclination to move toward using computer applications for managing company systems is one of the fundamental aspects in terms of measuring a company's ability to adapt to an e-Mall environment. Therefore, the questions in this section focus on the three main points. Firstly, the products sold by the companies, which indicates 72.7% of the companies sold products rather than services, which mean that the availability of auxiliaries, such as shipping and delivery companies, is an important factor for these companies in accessing the e-Mall environment. Secondly, the existence of an MIS Department in the sample companies, the result indicates the availability of an infrastructure for adopting e-Mall framework in the future. Lastly, the kinds of computers used, which indicate 90.9% of the participants responded that they were using personal computers. Also, 76.4% indicated they were using mainframe stations. This indicates that most of the companies participating have the basic ability and fundamental requirements to enter into the e-Mall field in the future.

**Table 3: Attributes of SMEs participants**

| Category of SMEs | No. | % |
|---|---|---|
| All participating | 110 | 100 |
| 1- Business type of the participant company | | |
| Real Estate and Public Services | 6 | 5.45% |
| Books/Learning Resources | 6 | 5.45% |
| Electronics and Electrical | 18 | 16.4% |
| Car parts and Accessories | 10 | 9.1% |
| Beauty Products | 10 | 9.1% |
| Clothing | 14 | 12.7% |
| Furniture | 10 | 9.1% |
| Jewellery | 8 | 7.28% |
| Medicines and Medical Supplies | 2 | 1.8% |
| Household Items, Tools, and Toys | 8 | 7.27% |
| Tickets, Hotel, Car Rental | 6 | 5.45% |
| Computer related | 6 | 5.45% |
| Grocery*- | 6 | 5.45% |
| 2- Using the Internet for buy or sell | | |
| Yes | 44 | 40% |
| No | 66 | 60% |
| 3- Company size | | |
| Less than 10 | 14 | 12.7% |
| 10–59 | 36 | 32.7% |
| 60–100 | 60 | 54.5% |
| 4- Product type | | |
| Intangible products | 6 | 5.5% |
| Tangible products | 80 | 72.7% |
| Both | 24 | 21.8% |
| 5- Existence of MIS Department | | |
| Yes | 90 | 81.8% |
| No | 12 | 10.9% |
| No Input | 8 | 7.3% |
| 6- Kinds of computers used | | |
| PC | 100 | 90.9% |
| Main Frame | 84 | 76.4% |
| Laptop | 48 | 43.6% |
| Mobile | 28 | 25.5% |
| 7- Internet Access Availability | | |
| Yes | 104 | 94.5% |
| No | 6 | 5.5% |
| 8- The Internet type | | |
| Broadband | 50 | 45.5% |
| DSL | 54 | 49.1% |
| Dial Up | 2 | 1.8% |
| Connect | 8 | 7.3% |
| Smart Phone | 6 | 5.5% |
| 9- Business type | | |
| Real Estate and Public Services | 6 | 5.45% |
| Books/Learning Resources | 6 | 5.45% |
| Electronics and Electrical | 18 | 16.4% |
| Car parts and Accessories | 10 | 9.1% |
| Beauty Products | 10 | 9.1% |
| Clothing | 14 | 12.7% |
| Furniture | 10 | 9.1% |
| Jewellery | 8 | 7.28% |
| Medicines and Medical Supplies | 2 | 1.8% |
| Household Items, Tools, and Toys | 8 | 7.27% |
| Tickets, Hotel, Car Rental | 6 | 5.45% |
| Computer related | 6 | 5.45% |
| Grocery*- | 6 | 5.45% |
| 10- Website | | |
| Business already has a website | 74 | 67.2% |
| Company plans to have a website | 8 | 7.3% |
| No website | 28 | 25.5% |
| 11- Purpose of the website | | |
| No identified purpose | 30 | 27.3% |
| Presenting product information | 60 | 54.5% |
| Customer service and support | 52 | 47.3% |
| Distribute offers and new products | 36 | 32.7% |
| Sale of part, or all, of company products | 26 | 23.6% |

The second part in the survey concentrates on the communication with customers via online means such as with e-mail which is one of the characteristics of successful companies in the modern era. Therefore, this part focuses to measure the companies' sense of how to implement these electronic channels and how these aspects can be used to achieve their vision. This part is divided into five questions: the availability of the Internet within the company, the Internet type, the existence of an MIS Department in the company, the company website and its purpose, and the product types sold online. For the first question, 94.5% of the participants indicated that there was Internet within the company, and most of the participants used the Internet with fast services, such as DSL and Broadband. This indicates that such firms are keen to provide Internet that is rapid and appropriate for their business, as they want their businesses to be as effective as possible. For the second question, 67.2% of the participants indicated that their companies have already websites, while another 7.3% planned to design websites in the future. This shows that companies are aware of the importance of Internet sites in promoting their business. The goal of website design was approached in question four, with four purposes. The largest group encompassed 54.5%, indicating participants who designed their website to present their product over the internet, and this is relevant to the previous question, which dealt with the importance of the Internet website in activating company commercial activity. Finally, question five investigates the products that can be sold over the Internet. The top five answers are as follows: 76.4% chose electrical and electronic products, 68.22% chose beauty and health products, 50.9% chose furniture, 43.6% chose clothing, and 41.8% chose automobiles.

### 4.2. The main findings for under consideration variables

The results of the questionnaire present certain aspects that affect the diffusion of worldwide e-Malls, and many of these aspects have an impact, to varying degrees, on Saudi Arabian society. These variables can be displayed according to the group factors, which have been presented in the literature review part.

#### 4.2.1. People and organizational variables

The previous figures show that the first group's results included eight variables, which have been numbered from 1 to 7 and 17th in Table 3. It is clear that the agreement average of community sample exceeded 82.2% of sellers, indicating these factors are influential in the diffusion of the e-Mall in Saudi Arabia. The percentage of those sampled who agree with the impact of education level on the e-Mall is 96.3%. Also, 85.1% of sellers agree that past experience has an effect on activating electronic markets. Furthermore, the company's size and its level of income is one of the most influential variables in the spread of electronic markets, with 67.6% of sellers approving this, and 75.9% of sellers also answering that the type of business activity influences the spread of e-Malls. This idea supports the results from the questionnaire on the type of products that can be sold or purchased over the Internet (see Q7 in Table 2), which includes the commercial products that people prefer to purchase over the Internet. Moreover, the aspirations of consumers and sellers are important factors affecting the spread of e-Malls, with 63% of sellers agreeing that the reality of current e-Malls does not meet the needs of consumers. Additionally, the benefits of e-commerce, and its impact on the spread of electronic markets globally, is one of the main aspects approved by the sample of community, 94% of sellers agreeing with this idea in the questionnaire results. This demonstrates the importance of this aspect in the spread of the e-Mall in Saudi Arabia. Finally, the weakness of the essential sources that are required to increase the IT skills necessary for e-procurement is one of the most fundamental and important factors effecting the spread of electronic markets, with 81.4% of sellers agreeing that this factor, and the lack of these skills and information sources, may have a negative impact on the diffusion of the e-Mall in Saudi Arabia. These factors, in relation to the survey results, show that most of the sample agrees on the impact of these factors, particularly in terms of Saudi society.

The fourth question deals with the participants' views concerning the influence of employee online trading awareness and education on the diffusion of e-Malls. This question has been asked to the sellers for two reasons: first, for understanding the sellers' opinion about this issue, as it relates to them directly, and should be an essential part of the company's activities; and second, because the e-commerce activities may not be an essential activity for the company, and it can be activated by an agent who working with company and he can deal with this matter behind the company's vision. Therefore, this should be asked only to seller participants. The results indicate that 96.3% of the participants agree on the importance of this aspect, and its impact on the spread of e-Malls for vendors. This means that the majority of vendors want to participate in e-Malls under their own supervision, rather than under third party control.

#### 4.2.2. Technological and environmental variables

The results for the technological and the environmental variables reveal some of the most important aspects of Saudi society's views on what impacts electronic markets in Saudi Arabia. Therefore, we believe these results indicate the willingness of Saudi society to adopt the e-Mall system. In this part, variables have been addressed that affect the infrastructure of the delivery system and the telecommunication channels, as well as the means that help and keep the websites safe and stable. There are six questions, dealing with the current infrastructure of online trading, postal services and delivery systems, secure personal computer operating systems and websites dealing with business and financial aspects, safe payment methods, activate secure and inexpensive payment means cooperating, and electronic web applications that are supported by the Arabic language.

First question about the impact of the Current Infrastructure on the online treading, 90.7% of vendor participants agreed with this opinion. The results also show. This means Part of the participants are content with the current communications infrastructure and do not believe it is influential for an applied e-Mall framework. Second, for the postal services and delivery systems variable, the questionnaire shows that 90.7% of the sample agrees on its importance for the spread of electronic markets. Third, 77.8% of the participants agree that the website should support the Arabic language. It is important to use the language of the community in activating the e-market environment in Saudi Arabia. These questions relate to the infrastructure, its readiness requirements and the basic needs for the activation e-Mall framework.

Table 4: The results of the numbers and percentages of each element in the questionnaire

| Variable No. | The Variables | Seller Count | Seller Rate |
|---|---|---|---|
| V1 | level of educational impact on the diffusion of the e-Mall | 104 | 96.3% |
| V2 | Impact of previous experience on activating the diffusion of the e-Mall | 92 | 85.1% |
| V3 | The company size and level of income, and their influence on activating the e-Mall | 73 | 67.6% |
| V4 | The level of online trading awareness and education in your corporation | 104 | 96.3% |
| V5 | The types of business activities that should be used via e-Mall | 82 | 75.9% |
| V6 | Lack of essential recourses and IT skills for online trading | 88 | 81.4% |
| V7 | Sites that use these systems are not compatible with consumer needs | 68 | 63% |
| V8 | Current Infrastructure of online treading | 98 | 90.7% |
| V9 | Activate postal services and delivery systems | 97 | 90.7% |
| V10 | Secure personal computer operating system and website dealing with business and financial aspects | 92 | 86.8% |
| V11 | Safe and protected payment methods | 102 | 94.5% |
| V12 | Activate secure and inexpensive payment means cooperating | 91 | 84.8% |
| V13 | Electronic web applications are supported by Arabic language | 84 | 77.8% |
| V14 | E-Malls are an alternative for women to visit the Mall centres | 43 | 39.8% |
| V15 | Promoting development through e-Malls for country areas away from major cities | 74 | 69.2% |
| V16 | E-Malls reduce the social benefits of shopping | 56 | 51.9% |
| V17 | E-Malls limit price negotiating between sellers and their customers | 80 | 74% |
| V18 | The importance of spreading the online shopping benefits | 100 | 94% |
| V19 | Your opinion about the importance of e-Mall framework system effectiveness in Saudi Arabia | 90 | 83.3% |
| V20 | Trust e-Mall framework will be helpful to increase the consumers' confidant to diffuse this system | 96 | 88.9% |
| I1 | The importance to deal with internet commerce for small and medium enterprise | 98 | 90.7% |
| I2 | The importance to provide an e-Mall framework for the portal for dealers who cannot create private Internet sites | 96 | 88.9% |
| I3 | The importance to Regulations and laws to protect the consumer and the seller over the Internet | 108 | 100% |
| I4 | The e-Malls system is provide employment opportunities and reduce the spread of unemployment | 68 | 63.6% |

Other technical aspects concentrate on the means to protect the computer and Internet sites from intrusions, as well as secure and cheap payment means. The following results is contained the target audience view. The first question focused on the security aspect, with 79.4% of the contributors agreeing that this aspect is important in activating electronic markets in Saudi Arabia. In terms of saving and securing means of payment over the Internet, 94.5% of the sample responded that this aspect is significantly important. The proportions may be high because the sellers will be receiving the money and the consumers will be paying the money, and they both want guarantees that the money is going to the right place. Results for the third question indicate, in terms of activating the secure and inexpensive means of payment, that and 85% of vendors sector agree it is important in activating electronic markets, as well as influencing the adoption of the e-Mall in Saudi Arabia.

These variables show the importance of the technical aspects, infrastructure, and a safe means of payment in activating electronic markets. The average for these variables was 87.3% for vendors group. This shows the extent of the role played by these factors in activating electronic markets globally and in Saudi Arabia.

*4.2.3. The third Group result: traditional and cultural variables*

In this part, the results address the impact of culture and tradition on the e-Mall in Saudi Arabia. The previous studies have shown six variables affecting this aspect, with the results presented in Figure 7. The following are the most significant findings.

Questions 14 and 16 focus on the social characteristics of the e-markets. These questions deal with whether the e-Mall could be an alternative, for women, to visiting the traditional markets, and whether the e-Malls reduce the social benefit of shopping. For the first question, 39.8% of the seller segment agreed. This indicates that, in terms of tradition, Saudi society does not consider e-Malls as an alternative to traditional Malls. Both of traditional Malls and electronic Malls have their audience, who are interested and keen on it. For the second question, 51.9% of participants agreed, but it is not a significant variable because both the traditional and e-Malls have their own special audience.

Questions 15, 17, 19 and 20 concentrate on the impact of cultural aspects. In looking at whether remote and country areas are the main beneficiaries of an e-Mall framework, 69.2% of participants agree. This indicates that this variable is an important influence on the e-Mall, as it can help target this group as an e-Mall community audience. In terms of limiting the negotiation between sellers and buyers, on the specific details and prices of products, 74% of venders group agreed this was important. Therefore, mechanisms that are helpful in keeping out negotiation between the seller and the buyer are beneficial, and this cultural aspect should not be a large obstacle in applying the e-Mall framework. In terms of being convinced about adopting this system in Saudi Arabia, 83.3% of contributors are convinced, and their personal conviction is that it would be helpful to let them use the system soon. In terms of whether public confidence will help with the application of this framework, 88.9% of participants agreed. They were convinced that trust is the key variable to a successful diffusion of this system in Saudi Arabia. Therefore, raising the confidence of the customers represents one of the biggest current challenges.

*4.2.4.    The main findings for the hypotheses part*

There are some assumptions that have been tested with a questionnaire about the applicability of e-malls in Saudi Arabia. The results came as follows

The first question deals with the importance of e-commerce for SME, with 90.7% of sellers agreeing; this is one of the encouraging findings in this section. The second question deals with the importance of regulations and laws to protect the customer during Internet transactions. The sellers agreed 100%. This demonstrates the fundamental importance of this aspect, especially in a bureaucratic country dealing with the new field of e-commerce. The third question deals with the importance of the government providing an e-Mall framework as a portal for dealers who cannot create private Internet sites, with access paid for as a membership fee, as well as whether having electronic transactions monitored by a neutral third party (the government) would give confidence to the consumers. The results indicate that 88.9% of traders, agreed that this aspect is important, which indicates the importance of this model, especially in Saudi Arabia, in terms of correctly organizing and establishing this framework. This model can later be implemented via companies in the private sector, similar to the Saudi stock market.

**4.3. Apply DOI**

After presenting the significant relationships between e-Mall its affecting variables, it should be concentrated on the relationships between DOI theory factors and how these relationships can be applied to the diffusion of e-Malls in Saudi Arabia. Therefore, it should start by focusing on two fundamental statistical tests that help us to ensure that these variables are suitable for analysing together and help us to examine the DOI factors clearly and accurately. These tests are the validity test (Cronbach's Alpha) and the multiple linear regression test. Below, there is brief explanation of each of these tests.

*4.3.1.    Reliability analysis test*

Cronbach's alpha is a measure of internal consistency, that is, how closely related a set of items are as a group. A high Cronbach's alpha indicates good internal consistency and suggests that the items measure one fairly similar idea [50]. Generally, a Cronbach's alpha greater than 0.60 is considered to meet the criteria of reliability. Use of Cronbach's alpha indicates whether each variable is good enough to use [42]. In our case, Cronbach's alpha was conducted on all variables and the result is 0.955. The sprite variables are as follow:

| Table 5: Data analysis from all respondents showing factors listed in order of descending mean value for each group ||||||||||
|---|---|---|---|---|---|---|---|---|---|
| | | One-Sample Test ||||||Cronbach's Alpha |||
| | | Test Value = 0 |||||| |||
| Variable | N | t | Std. Deviation | Sig. (2-tailed) | Mean Difference | 95% Confidence Interval of the Difference || Corrected Item-Total Correlation | Squared Multiple Correlation | Cronbach's Alpha if Item Deleted |
| | | | | | | Lower | Upper | | | |
| V1. | 109 | 35.289 | 1.27297 | .000 | 4.30275 | 4.0611 | 4.5444 | 0.847 | 0.885 | 0.951 |
| V2. | 109 | 34.740 | 1.16350 | .000 | 3.87156 | 3.6507 | 4.0925 | 0.714 | 0.827 | 0.953 |
| V3. | 109 | 30.451 | 1.22044 | .000 | 3.55963 | 3.3279 | 3.7913 | 0.471 | 0.725 | 0.955 |
| V4. | 109 | 40.647 | 1.02740 | .000 | 4.00000 | 3.8049 | 4.1951 | 0.735 | 0.713 | 0.953 |
| V5. | 109 | 36.377 | 1.09271 | .000 | 3.80734 | 3.5999 | 4.0148 | 0.706 | 0.792 | 0.953 |
| V6. | 109 | 34.623 | 1.18405 | .000 | 3.92661 | 3.7018 | 4.1514 | 0.787 | 0.877 | 0.952 |
| V7. | 109 | 33.821 | 1.09318 | .000 | 3.54128 | 3.3337 | 3.7488 | 0.428 | 0.594 | 0.955 |
| V8. | 109 | 34.979 | 1.22127 | .000 | 4.09174 | 3.8599 | 4.3236 | 0.674 | 0.819 | 0.953 |
| V9. | 109 | 37.074 | 1.15742 | .000 | 4.11009 | 3.8903 | 4.3298 | 0.811 | 0.858 | 0.952 |
| V10. | 108 | 33.491 | 1.29578 | .000 | 4.17593 | 3.9287 | 4.4231 | 0.824 | 0.926 | 0.951 |
| V11. | 109 | 35.787 | 1.24456 | .000 | 4.26606 | 4.0298 | 4.5023 | 0.845 | 0.879 | 0.951 |
| V12. | 108 | 32.191 | 1.30031 | .000 | 4.02778 | 3.7797 | 4.2758 | 0.783 | 0.779 | 0.952 |
| V13. | 109 | 34.774 | 1.22572 | .000 | 4.08257 | 3.8499 | 4.3153 | 0.752 | 0.744 | 0.952 |
| V14. | 109 | 24.639 | 1.30617 | .000 | 3.08257 | 2.8346 | 3.3306 | 0.324 | 0.695 | 0.957 |
| V15. | 109 | 28.115 | 1.34570 | .000 | 3.62385 | 3.3684 | 3.8793 | 0.641 | 0.872 | 0.953 |
| V16. | 109 | 35.766 | .99087 | .000 | 3.39450 | 3.2064 | 3.5826 | 0.187 | 0.594 | 0.957 |
| V17. | 109 | 32.760 | 1.20460 | .000 | 3.77982 | 3.5511 | 4.0085 | 0.459 | 0.673 | 0.955 |
| V18. | 108 | 37.530 | 1.13840 | .000 | 4.11111 | 3.8940 | 4.3283 | 0.815 | 0.857 | 0.952 |
| V19. | 109 | 35.542 | 1.18577 | .000 | 4.03670 | 3.8116 | 4.2618 | 0.817 | 0.860 | 0.951 |
| V20. | 109 | 35.613 | 1.22106 | .000 | 4.16514 | 3.9333 | 4.3970 | 0.703 | 0.802 | 0.953 |
| I1. | 109 | 33.768 | 1.22821 | .000 | 3.97248 | 3.7393 | 4.2057 | 0.707 | 0.843 | 0.953 |
| I2. | 109 | 30.827 | 1.37022 | .000 | 4.04587 | 3.7857 | 4.3060 | 0.737 | 0.773 | 0.952 |
| I3. | 109 | 35.449 | 1.29965 | .000 | 4.41284 | 4.1661 | 4.6596 | 0.906 | 0.957 | 0.950 |
| I4. | 108 | 32.787 | 1.10644 | .000 | 3.49074 | 3.2797 | 3.7018 | 0.338 | 0.602 | 0.956 |

The main objective in using the Cronbach's alpha measurement here is to study the reliability between variables to create new groups depending on parts of the DOI theory. The DOI theory contains five main perspectives, some of which have sub-perspectives. All of these aspects are measured by effecting variables, which were mentioned previously (see Tables 6). Each part of the DOI theory is measured by a variety of variables coming from the different (technical - organizational - cultural) groups. The Cronbach's alpha test was employed to determine the suitability of the entry of each variable in a particular group, and whether these variables have a positive or negative effect. If any of the variables give a negative result (less than 0.7), it will be excluded from a new group, which will measure DOI theory aspects. If a variable has a positive affect (0.7 or more), it will be accepted for use in a new group. To achieve our goal, this test was applied along with a multiple linear regression test.

*4.3.2. Multiple linear regression test*

Multiple regression (R) is defined as a statistical tool that allows the researcher to test how single or multiple independent variables are correlated with dependent variables. The test provides a score between 0.00 and 1.00. In this case, if the relation is 0.3 or more, the link is significant and it is acceptable; if it is less than 0.3, it is not acceptable and it will be ignored. When the relationship between the variables is calculated, the explanation can be used for making more accurate predictions about the reasons for and status of a given situation.

Many of these variables are shown to reconsider on how to deal with e-commerce approach with some of the community's characteristics that have emerged through a clear relationship between the variable in groups 1 and 3 or between group 2 and 3, which was shown the importance of some requirements that should be provided, such as the infrastructure and falls below the safe and secure payment methods and support the commercial websites by Arabic language. As well as, the importance of automation and expanding the commercial sites features which including some of community characteristics and traditions, such as activating the role of negotiation between the buyer and seller on price and product by electronic methods, also increasing the credibility about the products that displaying via presenting accurate and clear information about the goods that increase the ability of participants to make their decision and increase the level of confidence about the activation of e-Malls in the future.

## [5] DISCUSSION

The main categories in DOI theory is adopted on the seller aspect in e-Mall framework as following.

*5.1. Perceived Attributes of Innovation*

The seller is one of fundamental part of the e-commercial process, and this should be focused upon when applying the DOI theory to understand the problems they may have in accepting e-commerce and their expectations of the e-Mall framework. This also defines aspects that are not addressed in the buyer's section. Only important statistics related to the obstacles seller side face have been discussed, and the same theory is applied in the consumer segment in another publication.

*5.1.1. Relative Advantages*

The e-Mall provides different advantages to sellers. There are many features for this part, including the ability to easily reach a wider geographical segment, increase the margin of sales profit, use already established electronic distribution channels and cut down on costs ([1],[4],[12]). These features are also what attract vendors to distributing their products through e-Malls. The most important questions, however, are what are the most helpful means in increasing the awareness of the traders' community to the advantages of e-commerce and what would help them accept and contribute to this framework?

The questionnaire results show the same variables affecting the consumer sector and highlight the importance of measuring participants' satisfaction with essential features of e-Malls and the importance of activating this system in Saudi Arabia. According to research result, 98% of online vendors see the importance of both these aspects. Of the respondents, 95% are SMEs, 92% of them require a delivery system, while another 92% have existing websites and are interested in e-commerce.

**Table 6: Variable determining the rate of innovation adaption – consumer an seller sector**

| The Seller Sector | | | | |
|---|---|---|---|---|
| *Variable Determining the <u>Rate of Adoption</u>* | | *Seller (After)*** | *Reliability test* | *Liner regression test* |
| 1- Perceived attribute of innovation | | | | |
| | ▪ Relative advantage | V18, V19 | 0.740 | 0.397 |
| 2- Compatibility | | | | |
| | ☐ Social Cultural Value | V18 | 0.625 | --- * |
| | ☐ Previously Introduced | V1, V2, | 0.746 | 0.514 |
| | ☐ Client Needs | V8 , V9 , V10 , V11 , V12 , V13 , I3 | 0.702 | 0.632 |
| | ▪ Complexity | V9, V11, V12, V7, I2, I3 | 0.652 | 0.459 |
| | ▪ Trialability | V20 , Q12 | 0.951 | --- * |
| | ▪ Observability | Q1 , Q8 , Q9 | | |
| 3- Type of innovation decision | | | | |
| | ▪ Optional | V1, V2, V3, V4 ,V5 , V6, V19 , I1 , I2 | 0.683 | 0.397 |
| | ▪ Collective | | | |
| | ▪ Authority | | | |
| 4- Communication channels (e.g. mass media or intrapersonal) | | Q2 , Q3 , Q4 , Q5 , Q6 , Q7 | --- * | --- * |
| 5- Nature of the social system (e.g. its norms, degree of network, interconnection, etc.) | | V14 , V15 , V16 , V17 , I4 | 0.777 | 0.525 |
| 6- Extent of change agents' promotion efforts | | I1, V19, | 0.638 | 0.469 |
| * Multiple linear regression tests cannot be used at this stage because the variables here do not have matching characteristics. | | | | |

All these figures indicate that increasing people's awareness and knowledge of e-Malls in Saudi Arabia is essential at this point. Since most traders are usually ready to adopt a new trading style that helps them cut down production costs leading to additional profits, most of them would consider this framework.

The variable on previous experience of IT may not appear as important as the level of educating people about the importance of activating the new relative advantage to the seller segment. However, raising awareness about the advantages of e-commerce may be the biggest incentive to encourage sellers to participate in activating e-Malls.

### 5.1.2. Compatibility

Compatibility is a key point for the seller to lets them adopt or stop depending on their own views. Therefore, successful trade stems from a consumers' willingness to buy online, especially if buying online is clearly beneficial. Therefore, sellers should look for ways to make their systems compatible with the buyer by focusin g on the following aspects:

*Social and Cultural Values*

Understanding social and cultural values is important for traders if they want to be successful in a global e-commerce environment. Sales via the Internet are expanding both in local and foreign markets because traders are starting to understand the importance of getting involved in the e-market to meet consumers' needs. Consequently, traders should become more aware about the importance of getting involved in the e-Malls framework in the future. The aspect of cultural and social values can be summarized as follows:

A. Presenting high-quality products at competitive prices, which will increase the seller's reputation and make customers more confident in e-Malls [12].
B. Opting for e-sales, which is fast becoming a trend in most government services; in the survey of seller participants, 85% agreed with the effectiveness of e-business in Saudi Arabia. In addition, the rate of firms' readiness with online trading is 29.1%; and according to how convinced they are of the importance of the e-Mall system, this percentage will be increased soon.
C. Cashing in on the resulting benefits of e-commerce. This was confirmed by 98% of the company participants who already practice online selling.
D. Creating opportunities for companies that cannot participate in e-Malls and dealing with agents to distribute their products sold electronically. This means the dealer, distributor and the buyer will all gain through e-trading, which was agreed with by 63.6% of the company participants.

These points can increase the attractiveness of the e-Mall in the traders segment, and increase their convicted of the importance of online trading.

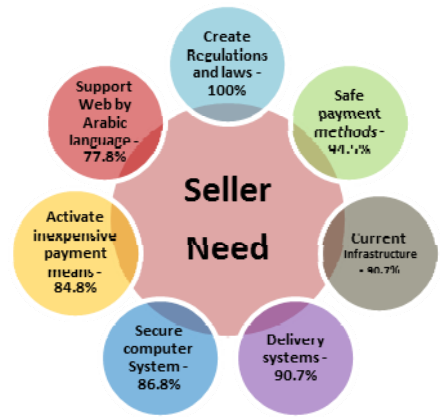

Figure 4: Seller needs in the e-Malls

*Previously Introduced*

Talking about already successful ideas is a key aspect in boosting traders' courage to participate in e-Malls. However, this experience can be fraught with some risk, especially since it is new for sellers in Saudi Arabia. Traders are often very concerned about studying the economic feasibility of any new project in which they might want to participate, such as the online marketplace. The feasibility of this market can be measured by using two basic variables: educational level and previous experience. The level of education provides the dealer with the basic knowledge that is needed to participate in this system [35], while previous experience provides the basic aspects of feasibility study [19]. Therefore, 94% of the companies that have never dealt with online sales place importance on educational level and 88% find previous experience in this aspect more important. Therefore, grouping participating traders together will allow them to support each other. New traders who wish to enter the online marketplace should consider the following:

A. Understanding the suitability of an electronic commercial activity is necessary for traders because it is one of the basic requirements needed to qualify for entry into the e-market which is presented in the next point client need.
B. The number of successful traders involved in the business activity.
C. Encouraging customers to buy online through some promotions. This was confirmed by 69.2% of the survey participants.

*Clients' Needs*

Meeting an e-client's needs is one of the key factors to active participation in an online marketplace. This was confirmed by 91% of the companies, who said in the questionnaire that they knew about online sales, but only 36% of them had actually conducted a sale via the Internet. Furthermore, only 60% of the companies have a website and only 40% of them aim to sell over the Internet. Therefore, though many of the actors have considered online trading, some do not have confidence in their ability to deal with e-sales. Hence, focusing on the actual needs of the client will help buyers enter this field with more confidence. The points that will help boost sellers' confidence in e-trading can be summarized as follows:

A. *Current Infrastructure of online treading:* The key to this aspect is providing the required infrastructure to expand e-Malls by activating fast Internet links between the companies' main information centres and their branches. In this area, the speed of the Internet (DSL or broadband) is not the only requirement. According to a report from the Ministry of Communications and Information Technology, companies that have branches around the Kingdom believe that the quality of broadband is not good enough to link up all their offices [14]. Therefore, a faster broadband speed is essential to improving the level of performance, as well as increasing the quality of service provided in Saudi Arabia's different cities. Of the companies surveyed, 90% said they were linked up to the Internet, with 51.9% of them using a DSL connection and 48.1% using broadband. However, only 40.07% of those that have broadband Internet access are selling online, which is a low percentage if it is measured against the companies that are interested in online trading.

B. *Activate delivery systems:* Activating delivery and the home mailbox system could help build an essential support system in the vendors' sector which is share with consumer concern and it is one of the major outcomes and external work of e-Mall system. Especially when we consider that the Internet is growing out in the mail system, so it should work to develop delivery systems to suit with customer requirements over the Internet. In the survey result, it was indicated that 91% of the participants who sold their products electronically needed this aspect to use e-Malls. In contrast, using private shipping companies to deliver goods sold online can have the negative effect of increasing costs.

C. *Secure computer operating system:* This was mentioned as one of the many reservations of sellers segment. The main reason is, most companies have a huge database, they need to consider how best to protect it. There were mentioned by 74% of the companies in the survey sample. Furthermore, e-commerce requires support programs by production companies in the form of original software, which increases the awareness level and concern in these companies.

D. *Protected payment methods:* This aspect is more important for traders than buyers. After every transaction, a contract is made between the company and the electronic payment gateway to receive the money from the buyer and send it to the dealer. Up to 91% of the companies trading online agreed with this. These gateways differ in their requirements and the degree of their performance. Their suitability can be measured according to four key points:
- Dealings in the Middle East and Gulf countries in particular;
- The means of checking on credit-card validation and data;
- The documents and fees required to open the account;
- Additional features to assist customers such as transferring money to their client's bank account.

A comparison has been done between popular e-payment gateways and the advantages they offer to each company (see Appendix A).

E. *Activate inexpensive payment means:* The seller does not want to sell high-cost products if he can help cut down the cost or cancel some fees. In the survey, 88% who agreed upon the importance of this aspect were SMEs, who said that because they have limited profit, they do not need to increase their costs by paying unnecessary fees, especially if they want to attract customers via electronic channels. Therefore, dealing with global payment gateways that can provide relatively good promotions to vendors and e-Malls will be reflected in customer satisfaction and a reduction in procurement costs.

F. *Support e-commerce in the Arabic language:* This is very important because it creates the appropriate environment to display the products in full detail, which is helpful to increasing consumer confidence gradually. Besides that, it is beneficial for the seller to present details about their products. This aspect was agreed upon by 78% of the companies that have websites.

G. *Create regulations and laws that protect e-traders:* This aspect is important because traders need to make investments in order to increase their profits. However, they cannot make this decision if all the procedures are not systematic and legal. All of the participants in the questionnaire highlighted the importance of implementing a legal system that protects them as well as the buyer in all e- transactions. As is well known, these regulations help limit unauthorized vendors from cheating customers by directing them to unsafe sites on the Internet. Furthermore, the existence of these laws will help increase the traders' confidence in e-marketing [34]. The authorities should also activate policies that ensure readiness on the part of both the sellers and the buyers to enter into transactions in the online marketplace. These policies will help both the vendor and the consumer gain trust in the e-Mall procedure.

The difference between a seller and a buyer dealing with online marketplace regulations is that the buyer can completely end his relationship with the transaction upon receipt of the goods. In case a problem arises between the seller and the buyer, a strict management system can help resolve these problems. However, the vendor's connection with the e-Mall will continue for a long time; therefore, it should to activate a government oversight role to help create a reassuring environment of investment for dealers. This regulation can be formulated in collaboration with the Commerce and Labour Ministries by drafting these regulations, which can be approved by the authorities and relevant institutions to create a safe environment for e-trade.

5.1.3. **Complexity**

Many systems should be activated for sellers in e-Malls, which were represented by the coefficient statistical correlation with each other. These systems can be divided into six sub-systems as follows:

A. Complexity related to the home-delivery system: E-commerce faces many challenges in Saudi Arabia that have been mentioned earlier in the consumer section. The seller wants to deliver the goods to the buyer otherwise it will be obliged to return the amount paid by the buyer, which will cost the seller extra in chargeback fees. This aspect was concerned to 90% of survey participants of SMEs. If the delivery system is not at a sufficient level of efficiency, it will further undermine confidence between the seller and the buyer, which is essentially needed to make the e-Mall framework work.

B. Electronic payment gateways: These channels can be subscribed to by trader, or let the e-Mall management subscribe to this part of the system for all members of the e-Mall. 97% of survey sample, who are not move toward the online trading, believe the importance of this aspect. Therefore, it should be provided local or regional e-payment gateways that can provide security in exchange rates and competitor for foreign payment gates.

C. Many traders prefer local payment gateways such as SADAD, which offers free payment options for consumers and is a cheap choice for sellers. 86% of participants companies who practice of online sale were believed importance of this side. Therefore, since traders aim for cheap payment means, it is becoming more important to set up and activate local or regional e-payment gateways in the near future.

D. Measuring the compatibility of the site with the wishes of the customer: This is a very important variable in successfully managing an e-Mall [37]. The first issue is the ease of use and compatibility of an e-commerce site with the desires of the target audience that help raise the level of performance and increase the number of visitors to the website.. The importance of this aspect was confirmed by 64% of the participants, companies that sell products as their main activity, as well as 67% of those who sell services.

E. Government support is important because most traders feel that dealing with new ideas on their own is far too risky, so many prefer government regulations even if it is just logistical support. Many traders believe that having the e-Mall under a government portal would help increase consumer confidence and help the seller flourish in a low-cost business environment. However, having the market portal under government regulations may not be attractive to all traders because some may not be too concerned about customer satisfaction or making profits, especially if these gains are not reflected in the performance of government-sector employees [34]. However, about 91% of the companies that do not sell on the Internet believe that e-markets need government support, especially in helping companies survive in a competitive environment, helping create an e-market system that is subsidized by government and is also compatible with the customers' wishes, so buyers can be confident and satisfied.

F. Despite the importance of the government regulations on e-Malls, which will have a significant effect on the performance of the business, setting up a system complete with its hierarchy might take several years [34]. In order to set up and implement laws of this nature, many authorities need to be involved, such as the Justice, Labour, Commerce, Communications and Information Technology Ministries. Therefore, addressing this matter at the moment can be too difficult and complex. It is very useful the participants' companies in the survey were coincided with the importance of regulations and judicial legislation in as one of the basic requirements of suitable investment in the e-Mall system and their rate was 100%, and there are 16 companies have different posts in the open questions, and they are agree with this side. For their examples are six of them said the importance law to protect both consumer and seller, eight responses about the importance of existence sides to except the online customer complaints, and two responses about the importance of computable the e-commerce system polices with the regulations of the Ministry of Justice. Therefore, perhaps the easiest way to initiate this issue quickly would be to activate a supervisory role in the online marketplace from the very beginning by drafting regulations before they are sent to the stakeholders. If this method is followed, then the draft can be written by people who understand online trading and what sellers and buyers need in the online marketplace, thus designing regulations for an e-Mall system will be faster and more realistic.

### 5.1.4. Trialability

The ability for consumers to try products is one factor that can help increase chance of purchase. At this point in the development of e-commerce, sellers are considering the types of electronic alternatives to provide to make product trial possible. It is important to consider what assistance means in order to continue to increase the uptake of experimental methods to enable product trialability, and the increase in consumer confidence that can result. Therefore, the open questions part contented 10 samples responses about the importance to try to view and display the product before purchase, also, it should be made payment upon delivery as propose from some participants some important, as well as the possibility to return the product to the seller, if it is not similar to specifications. These answers show the importance to consider in this aspect, to give the target population to trust the seller and e-Malls models through apply these methods.

The ability to segment the e-Market is part of trialability, in that it helps identify the target population in the online marketplace rapidly. A great need for consumers, a so-called 'segmentation e-Market' may help provide a suitable environment for the seller to view the products purchased according to the extent to which they were considered desirable by the targeted consumer segment. The vendors' willingness is represented by the affirmation that 83.5% of the vendors agreed to the importance of this aspect, is also confirmed by the convictions of consumers regarding the subject. Ultimately, the diffusion of e-Markets will be greatly facilitated by vendors being able to operate in an environment in which the products are displayed to their suitable target audiences in a fashion that allows trial before purchase in the Saudi e-Mall.

### 5.1.5. Observability

Observability refers to the ability to keep accurate statistics with respect to number of purchases, consumer characteristics, and other elements for the parties (vendors and consumers) operating in the Saudi e-Mall. The concept of the e-Mall should also lead companies to take steps to increase confidence among their consumers. Companies, for example, should have an existing, active Website. On this Website, the company's achievements should be mentioned without exaggeration, giving a real measure of the company's performance to consumers. Therefore, more than twenty responses in the open questions in the survey were focused on the importance of credibility in electronic sales. They were pointed to the importance of credibility in the handling and quality, between sellers with buyers, and in the presenting the product information in the websites. Companies should also present new products that will attract consumers to their sites again. Finally, companies should have communications on their Websites that deliver the newest product news to their target audience.

### 5.2. Types of Innovation Decisions

The following factors should serve to help companies make appropriate decisions to build the e-Mall:

- The education level and previous experience: 93% of SMEs believed in the importance of this aspect, as did almost all companies that solely sell products online.
- Company Brand Name: The brand name, company size and profit, and product franchising are helpful variables that can help companies bolster confidence in specific products among consumers [38]. Therefore, 73% of participants' companies who practice online selling consider with this aspect. Even though the online e-Mall is open to all companies to present their products, the brand name provides better-recognized companies with leverage among consumers. Thus, other companies lacking a strong 'brand' should search for new competitive standards that are compatible with their consumers' wishes; this will help them gain entry in e-commerce.
- The level of knowledge and awareness of company staff in executing online business: 100% of corporate participants believed in the importance of having in-house knowledge of e-Markets and their characteristics, as did 96% of SMEs. The primary reason for the importance of in-house knowledge is that outside vendors and overseas firms may not have a firm grasp on a company's overall business model. Besides that, adopting new model will cost the companies a lot to hire expert staff to deal with computer issues, and to develop the company's infrastructure to online trading [53], this is considerably diminishing the value of their expertise to adopt e-Malls system.
- Assistance sources and technical support: 80% of the companies that have information stored electronically pay attention to this aspect for several reasons, including the need to maintain a high level of information security when entering the electronic sales and purchases arena [18], and the need for a corresponding high level of information availability to all appropriate branches of a business to provide sufficient execution capability. Data must be dealt with very carefully to avoid hacking and to enable competing in the current business environment.
- The company's support to apply the e-Mall system in Saudi Arabia: To the seller, the consumer's desire is the first priority. In order to, the buyer in Saudi Arabia is turning to e-procurement for many benefits. Companies, accordingly, must showcase their products to the appropriate target audiences over the Internet. Therefore, 83.3% of participants' companies agreed to the importance of this side. It has been mentioned that consumer segmentation with e-Markets differs from consumer segmentation in traditional markets; this signifies that traders should be well-situated if they have a strong position in both e-markets and traditional markets.
- The importance of e-Malls for SMEs: Importantly, the e-Malls option must provide options to SMEs, which represent 95% of the businesses in Saudi Arabia. Therefore, the entry of SMEs into online commerce is inevitable sooner or later. Also, it was agreed by 90.7% of firms' participants in the survey. As a consequence, this sector should be supported and provided for with respect to facilitating the ease of uptake of the e-Mall model. The Boards of Directors at SMEs should also make the important decision to participate, which would be beneficial to e-Malls in Saudi Arabia.
- Government support through low-cost governmental online Mall portals: This will help provide unlimited support for SMEs, who were representing 86% of the survey participants, as they ramp up their online sales capabilities.

These aspects give online traders to make their decision and think broadly about the fundamental aspects needed for a thriving e-commerce presence in Saudi Arabia.

### 5.3. Communication Channels

The traders segment that endorses e-markets is very wide, and the scope of this community can be summed up as follows:

- Companies that have an MIS Department represent 81.8% of participants; it can be focused on the medium-sized enterprises with an MIS department which represent 96.6% of participants, because these segment have a stronger financial capability in comparison with small and macro enterprises, and also have the readiness to store data electronically that should be necessary for e-Malls.
- The companies that have business in the favorite consumer segment is also important to consider; the different product segments for targeted consumer groups include electrical and electronic devices,

- clothing, beauty and health products, furniture, and automotive.
- Companies interested in high-speed Internet access (such as DSL or Broadband) represent 94.6% of the participants.
- Companies that have Web sites represent 67.2% of the companies of the study sample. There were 29.7% of these companies that already practiced online selling via their electronic Websites.

These groups are interested to selling online, which could be invited to participate in e-markets.

*5.4. Nature of the Social System*

According to many previous studies on Saudi Arabia and in the region in general ([4],[56]), social and cultural factors are critical factors in commercial transactions. To a large extent, the same responses regarding the nature of the social system were given by both vendors and consumers in our questionnaires. Therefore, it becomes clear that it is important to understand the nature of society, and think of the cultural values of those who live in a particular society, in order to develop appropriate business methods (either in an e-commerce or in a traditional context). Cultural and social factors that influence business processes and operations should be well understood, and the factors that primarily affect the buyer or the seller should be separately accounted for, especially since sellers attempt to follow consumers' wishes in general. The cultural and social factors divided into five key categories; more detail about them is given here.

1- *E-Malls provide an alternative to traditional markets for women:* No encouraging results exist in our questionnaire for vendors on this factor, as those with businesses are unsure whether e-commerce will be an appropriate alternative for women. 41% of the participating companies engaged in electronic sales agreed with the importance of this aspect; 33.5% of SMEs assented to its importance. In this particular case, these results confirm that vendors conform to consumer wishes and to what is needed to increase their customer base.

2- *Promoting e-Malls for non-urban areas:* 64% of those selling goods and services online, 71% of SMEs, and 75% of companies that just sell physical products agreed with the potential of e-commerce in this arena. Therefore, the non-urban consumer should be a target audience, and offers should be targeted to this population.

3- *The social benefits of shopping:* This factor is heavily weighted toward the consumer sector; that is, it does not represent something intrinsically important to the vendor sector. In the survey, 41% of practitioners of online sales agreed to the importance of this aspect, as did 46% of SMEs and 54% of companies selling physical products. A consumer segment exists that desires to shop via traditional markets, and one exists for online shopping. This factor therefore is not hugely influential in the minds of vendors.

4- *Limiting price negotiations between sellers and customers:* Negotiating (both its nature, and its results) counts as an important 'social values' element to most businesses [4]. 68% of businesses selling products via the Internet asserted its importance, as did a higher percentage (78%) of businesses that were not selling online. These percentages were fairly consistent across all types of businesses (72% for SMEs; 72% for companies with their own Websites). Ultimately, coping with negotiations in e-markets as part of the purchasing process is related to increasing the efficacy of communications channels between sellers and buyers; this can only facilitate the ease of electronic transactions and greatly enrich the online marketplace.

5- *The importance of e-Malls in reducing unemployment:* The unemployment rate in Saudi Arabia reached 43% in 2009 among young people, and 10.5% in the general population in 2010. Various opinions exist concerning whether e-Malls and e-commerce has the potential to improve the employment market, or cause further deterioration. People who believe e-Markets are helpful for reducing unemployment represent approximately 73% of the vendors who are not currently selling products online, and 50% of businesses that are conducting online business. 70% of companies with an MIS department perceive the importance of increasing their products' online sales, and reducing unemployment through this increase in business. On the other hand, many companies do not perceive this link; they believe they will lose many high- and low-level employees if they move to online business, and believe that e-commerce and e-markets will increase the unemployment rate. 50% of small enterprises and 68% of companies that do not have an MIS department have the same opinion. With this even split in opinion, vendors should reflect on the national duty all should feel to try and avoid massive disruptions of the job market; this is especially true of companies that currently do not have an online presence. These companies could boost employment by having an online presence in additional to physical locations. By either selling their products themselves or using intermediaries, these companies could employ those with no job and increase profit simultaneously, causing gains for everyone.

*5.5. Extent of Promotional Efforts on the Part of Change Agents*

Change agents must make all appropriate efforts to promote the adoption of e-markets in the region if they wish to see their desired change become reality. Providing an alternative to many of society's segments is a laudable goal, and persuading government agencies and others to adopt e-markets and make use of them could quickly increase uptake of e-Malls quickly. However, the main role of any promotional efforts would be to help companies develop a strong online presence to increase the group of online consumers, and create a nucleus of strong e-markets both locally and regionally.

*5.6. Summary of seller requirements*

The requirements of the vendor sector with respect to application of the DOI theory can be divided into five stages as already mentioned in the five stages of decision innovation process. The stages requirements as follows

In the knowledge stage, two key aspects are demonstrated in the seller sector: to increase seller awareness of e-market benefits and technical client needs. First, increasing e-commerce awareness of e-markets helps to raise the level of enlightenment in the target audience and audience satisfaction with respect to two aspects: to provide new advantages and its social value for e-Mall and assistance in identifying target aspirations of consumer and electronic needs. Second, the importance to determine e-Mall client needs for seller in the implementation of its online business. This involves seven elements, which are highlighted in the compatibility – client needs section.

The persuasion stage involves two key elements for the seller. These include increasing seller confidence in the e-Market and limiting the impact of cultural and social factors on e-procurement. The first aspect involves four subsidiary activities: applying legislation and policies governing the e-market, designing a trading website based on the expectations of consumers, previous experience in e-procurement and sales, and obviously publicizing the seller's achievements in the online trading field honestly. The other aspect is the cultural and social factors that affect e-procurement, which should focus on two important features. One is expansion of presentation methods for both city residents and those in remote villages. Another is expansion of basic ways to view the product clearly that help to increase confidence in approving the purchase.

The decision stage comes after the conviction stage. This stage has two main aspects: company readiness for online trading, and assistance resources and technical support for the e-procurement process and beyond. After that the implementation stage has the communication channels to reach the target audience easily and successively. The last stage is the confirmation which can be measured via increase the number of people who use the e-Mall in the future.

## [6] RECOMMENDATION

The important requirements that emerge from the seller's perspective can be summarized as follows:

A. As in the case of the consumer, sufficient technical expertise must be available to the seller. It is important to increase the knowledge of the technical requirements needed of the seller, which should be provided and activated within the e-Mall framework to fulfill the needs for cooperation between traders and e-Mall management.

B. Increasing the seller's confidence in several ways:
- Providing or encouraging government support.
- Designing e-markets based on consumer desires that can be used to create a suitable environment for attracting customers in the future.
- Highlighting the traders' successes through feedback programs, which help to increase consumer confidence.

C. The company's management and relevant departments should have the electronic commercial ability and skills to deal with the e-commerce sector effectively.

D. Meeting organizational and technical seller needs to bring investment into the online marketplace.

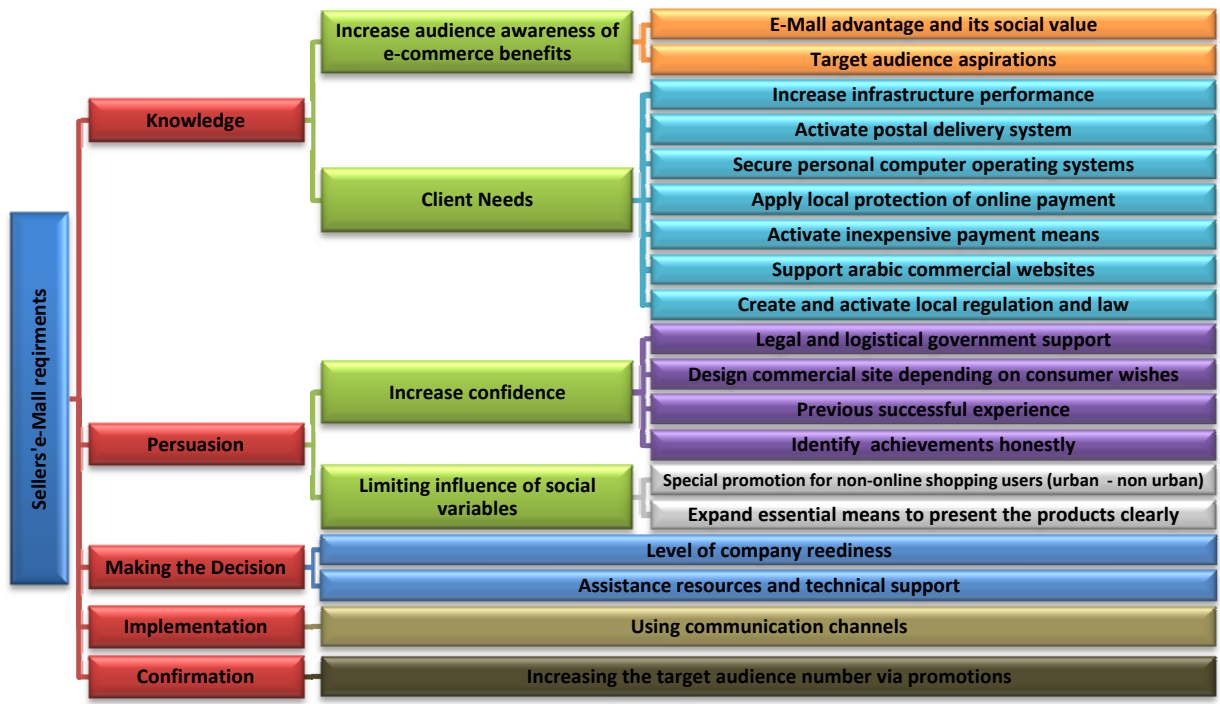

**Figure 5: Summary of e-Mall requirements of Seller**

Adel A. Bahaddad et al, / (IJCSIT) International Journal of Computer Science and Information Technologies, Vol. 5 (4), 2014, 5835-5856**Barriers**

Some of the obstacles faced by e-Malls activation represent another side of the research questions that has been demonstrated through the research outcomes. On the buyer side, some fundamental barriers discourage customers from buying and visiting e-Market sites as frequently as sellers would like. These shortcomings can lead to partial or complete reluctance to buy from e-Malls. There are many obstacles that have been appeared from the sellers' side, to activating e-markets too. There are some obstacles are shared by both sellers and buyers, which are divided into six key aspects and can be summarized as follows:

A. The confidence of the consumer in e-Malls, which is reported as one of major requirements of the consumer segment, needs to be increased. Therefore, the lack of availability of technical client needs will lead to the deterioration of consumer confidence with e-Malls, which will be a direct obstacle to adopting the use of e-Malls.

B. The impact of cultural and social habits on the adoption of e-Malls diffusion should be reduced. Two points are relevant here. Firstly, there should be fewer limitations on the ways venders and consumers can debate the details and prices of goods. Also, work should be done to utilize all of the various communication channels, which are suitable with all target segments of consumers' wishes to provide compatible communication environment with e-Malls.

C. There is a lack of government regulation dealing with e-Malls. These regulations are helpful in providing a safe environment to invest in this area.

D. There is no clear time plan for implementing the sellers' needs regarding e-Malls at the moment. These needs represent a major challenge in perfecting the adoption and use of e-Malls. Therefore, it is necessary to search for alternative options to activate e-Malls with reference to suitable foreign systems which are compatible with the e-Malls requirements in Saudi Arabia, such as e-payment gateways and other basic technical elements. In fact, these alternatives cannot be considered as the most appropriate options in online trading transactions as mentioned in previously, but they are suitable temporary alternatives to assist in establishing the use of e-Malls in Saudi Arabia.

E. There is a lack of assistance programs which works as decision support systems to display products realistically, which is one of the important requirements for consumers. It is necessary to work on providing the appropriate electronic environment to display goods properly and in full detail, to reduce the amount of negotiation on the details of the product and its price in the e-Malls. This will be helpful in decision-making, make purchasing easy, and save time. This issue should be addressed in cooperation between the venders and e-Mall management.

F. No clear means exists to attract consumers in remote areas and villages, although this segment is one of the important potential e-Mall markets.

## [7] CONCLUSION

One of the fundamental e-commerce entities in Saudi Arabia is the business sector. Although companies in Saudi Arabia have different levels and types of business activities, many large companies have established private commercial sites because of their financial ability to invest in the online trading [14]. However, this investment cannot be widespread because this segment of companies represents less than 5% of total companies number in Saudi Arabia, and its target audience is other smaller companies or government agencies [1]; thus, this trade type will be B2B and B2G, and the end consumer is not the target in this development. As a consequence, SMEs should not be isolated from this process, as they represent one of the cornerstones in the Saudi economy because of their dramatic ability to ensure the success of this approach. Such companies represent 95% of all of firms in Saudi Arabia ([6],[13]). Furthermore, they usually communicate with the final consumer, as demonstrated in many previous studies ([55],[48]). In addition, in many countries, it was this segment that entered into the e-commerce field, and such businesses have had a significant impact on the success the economy and online trade in those countries.

Many large companies have enough funds to move their business online, but SMEs may not have the financial capability to invest in this way. This requires activating several sub-approaches to e-commerce systems to allow SMEs to invest in a commercial website. This will give this segment a promising opportunity to invest in online trading over a wide geographical area. This is the main reason for studying the diffusion of e-Malls for SMEs in Saudi Arabia.

There is not enough e-commerce experience among SMEs in Saudi Arabia, which will impact negatively on online Saudi markets' ability to adopt the role of e-commerce. Therefore, understanding the requirements and difficulties related to factors influencing e-commerce on a global level, and examining them in Saudi, will help to identify the factors involved in activating the e-market in Saudi Arabia. If the e-Mall approach is not acceptable for the target population, this will represent a large obstacle. The seller will then be reluctant to deal with e-trade and discover its advantages. Therefore, the obstacles facing the implementation the e-commerce system from the perspective of society's opinion should be made clear. In addition, the actual requirements that help in the adoption of this system should be considered. Attending to both of these issues will help to forge a suitable path for the significant spread of e-commerce and online trading between businesses and consumers, particularly through e-Malls.

### REFERENCES

[1]. AboZaid, T. (2005). *Tishreen University Journal for Studies and Scientific Research- Economic and Legal Science Series* Vol. (27) No (4) 2005

[2]. Alfuraih, S. (2008). E-commerce and E-commerce Fraud in Saudi Arabia: A Case Study.*In 2nd International Conference on Information Security and Assurance* (pp. 176-80). Busan, Korea.

[3]. AlGhamdi, R., Drew, S., & Al-Ghaith, W. (2011). Factors Influencing Retailers in Saudi Arabia to Adoption of Online Sales Systems: a qualitative analysis. *Electronic Journal of Information*

www.ijcsit.com5854

### Appendix A: Comparison between global e-Payment gateways

|  | **PayPal** | **Alert Pay** | **Money Bookers** | **2Check Out** | **Plimus** |
|---|---|---|---|---|---|
| **Subscription fees** | Free | Free | Free | $49 | Free |
| **Arab countries supported** | All | All | All | All | All |
| **Arab countries have special privileges** | None | None | Bahrain & Kuwait | None | None |
| **Recharge balance by credit cards** | Free | Free | 1.9% | n/a | n/a |
| **Recharge the account by Arabic bank transfer** | n/a | $20 | Free | n/a | n/a |
| **Recharge the account by USA bank transfer** | Free | Free | Free | n/a | n/a |
| **Transfer fees to Arabic Bank account** | n/a | $15 | $2.41 | $15 | $21 |
| **Transfer fees to USA Bank account** | Free | $0.5 | $2.64 | Free | Free |
| **Fee per Transaction (Monthly)** | 3.4% + $0.3 | 5% + $0.25 | 2.9% + €0.29 | 5.5% + $0.45 | 15% for less 1000 4.5% $for more $1000 |
| **Less amount to be withdrawn** | $ 0.01 | $ 40 | $ 0.01 | $ 300 | $ 35 |
| *Source:* ([39],[44],[45],[62]) | | | | | |